\documentclass[twoside,fleqn]{ActaStyle}
\usepackage{times,cite}

\def\authorheading{ }
\def\shorttitle{ }

\begin{document}

\pagerange{1}{4}

\title{On the anomalous acceleration in the solar system}

\author{D.~Palle\email{palle@mefisto.irb.hr}}
{Zavod za teorijsku fiziku, 
Institut Rugjer Bo\v skovi\'c \\ 
Po\v st. Pret. 180, HR-10002 Zagreb, CROATIA}

\day{February 17, 2005}

\abstract{
We study an impact of the cosmological environment on the solar gravitational
system by the imbedding formalism of Gautreau. It turns out that the cosmic mean-mass
density and the cosmological constant give negligibly small contribution to the
gravity potentials. On the other hand, the cosmic acceleration beyond the 
Robertson-Walker geometry can considerably influence the curvature of spacetime in 
the solar system. The resulting anomalous constant acceleration towards the Sun
is order of magnitude smaller than that measured by Pioneer 10 and 11.
However, it is larger than the second order terms of potentials, thus well within the sensitivity
of new gravity probes such as the LATOR mission.
}


The general relativity (GR) corrections to Newtonian laws of planetary orbits are small
but included in many body dynamical calculations. Due to the great advances in
theoretical calculations by numerical codes and permanent technological 
achievements and improvements, numerous attempts are planned to measure
tiny corrections to even higher accuracies as a check of the theory of gravity.
In this paper we want to emphasize the importance of including some 
cosmological corrections to metric potentials and investigating whether they are
observable and distinguishable from the higher order parametrized post-Newtonian (PPN) terms.

We are faced with a problem of imbedding a Schwarzschild mass (in our case the
solar mass) into cosmological fluid. The approach of Einstein and Straus
is to cut out a spherical vacuum region and to put therein a Schwarzschild mass,
and then to join smoothly two different geometries at the boundary
surface. This approach looks quite unnatural because it excludes the
expected overlapping of gravity forces. Therefore, we choose a different approach
which allows merging of gravity fields, namely, that of Gautreau \cite{Gautreau}.
He relies on the introduction of geodesic time coordinates, which
makes possible defining cosmic fluid sources and imbedding 
of the Schwarzschild mass.

We deal with the so called flat (zero curvature) cosmic spacetime geometry,
favoured by current observations.
In adition to expansion, we include acceleration into the line element.

Small amounts of acceleration or vorticity in cosmological models
are allowed by current cosmological data. 
Moreover, recent claims on the large-scale asymmetry deduced from 
WMAP data could be explained by the presence of the vorticity (rotation)
of the Universe \cite{Palle1}.
On the other hand, even a small amount of acceleration, contributing to the 
cosmic geometry beyond that of Robertson-Walker, can 
strongly influence the evolution of the cosmic mass density contrast
at small redshifts \cite{Palle2}.
In the Einstein-Cartan cosmology with a spinning cold dark matter
particle one can derive relationship between Hubble's
expansion ($H_{0}$), vorticity ($\omega_{0}$) and acceleration ($\Sigma$) parameters owing
to the additional algebraic relation between the torsion 
of spacetime and the spin of matter \cite{Palle3}:

\begin{eqnarray}
 \Sigma H_{0}=\omega_{0}\frac{2}{\sqrt{3}},\ \  \Sigma={\cal O}(10^{-3}) . 
\end{eqnarray}

In addition, the Einstein-Cartan nonsingular cosmology can solve 
the problem of the primordial mass density contrast \cite{Palle4} and
the flatness problem \cite{Palle3}.

Let us neglect the vorticity and assume a flat cosmic geometry with
acceleration:

\begin{eqnarray}
ds^{2}=dt^{2}-S^{2}(t)[(1-\Sigma)dr^{2}+r^{2}(d\theta^{2}+sin^{2}\theta d\phi^{2})]
-2\sqrt{\Sigma}S(t)drdt, 
\end{eqnarray}
\begin{eqnarray*}
w^{\mu}\equiv u_{\nu}\nabla^{\nu}u^{\mu}\equiv\dot{u^{\mu}},\ w^{\mu}w_{\mu}=-\Sigma (\frac{\dot{S}}{S})^{2},
\\
u^{\mu}=velocity\ vector,\ w^{\mu}=acceleration\ vector . 
\end{eqnarray*}

The first task is to make a transition to the coordinates
where the Gautreau imbedding formalism could be applied.
We perform a general coordinate transformation (GCT) 
to the physical radial coordinate R=rS(t):

\begin{eqnarray}
g_{tt}=1+2\sqrt{\Sigma}\frac{\dot{S}}{S}R+(-1+\Sigma)(\frac{\dot{S}}{S})^{2}R^{2},\ 
g_{tR}=-\sqrt{\Sigma}+(1-\Sigma)\frac{\dot{S}}{S}R, \\
g_{RR}=-1+\Sigma. \hspace{80 mm} \nonumber
\end{eqnarray}

To reach a diagonal form of the curvature coordinates \cite{Gautreau},
we apply next GCT to the preceding metric \cite{Weinberg}:

\begin{eqnarray}
ds^{2}=C(t,R)dt^{2}-D(t,R)dr^{2}-2E(t,R)dtdR-R^{2}(d\theta^{2}+sin^{2}\theta d\phi^{2}), \\
dt^{\prime}=\eta (t,R)[C(t,R)dt-E(t,R)dR], \hspace{60 mm} \nonumber\\
\Longrightarrow ds^{2}=\eta^{-2}C^{-1}dt^{\prime 2}-(D+C^{-1}E^{2})dR^{2}-R^{2}d\Omega^{2}, 
\nonumber \\
condition:\ \frac{\partial}{\partial R}[\eta (t,R)C(t,R)]
=-\frac{\partial}{\partial t}[\eta (t,R)E(t,R)]. \nonumber
\end{eqnarray}

The functions C(t,R), D(t,R) and E(t,R) are defined directly by Eqs. (3) and (4) and
they are displayed below. Now we can
find $\eta$ function by solving its condition having a perfect differential $dt^{\prime}$
in power series expansion in the small variable $R H_{0}={\cal O}(10^{-14})$,
$H_{0}=H(today),\ H\equiv \frac{\dot{S}}{S}$:

\begin{eqnarray}
C(t,R) = 1+2\sqrt{\Sigma}HR+(-1+\Sigma)H^{2}R^{2},\ D(t,R)=1-\Sigma, \nonumber \\
E(t,R) = \sqrt{\Sigma}-(1-\Sigma)HR, \nonumber \\
\eta (t,R)=\eta_{0}(1-2\sqrt{\Sigma}HR+(1+\Sigma-\frac{q+1}{2}(1+\Sigma))
H^{2}R^{2}+ ... ), \\
\frac{\ddot{S}}{S}\equiv -q(\frac{\dot{S}}{S})^{2}. \hspace{50 mm} \nonumber
\end{eqnarray}

The constant $\eta_{0}$ can be removed by a redefinition of the time coordinate
$T\equiv \eta_{0}^{-1}t^{\prime}$(we keep only the terms linear in HR in
the curvature coordinates (T,R) form of the metric):

\begin{eqnarray}
ds^{2}=B(T,R)dT^{2}-A^{-1}(T,R)dR^{2}-R^{2}d\Omega^{2}, \nonumber \\
A=(D+C^{-1}E^{2})^{-1},\ B=\eta^{-2}C^{-1}. \hspace{20 mm}
\end{eqnarray}

Gautreau introduces geodesic time ($\tau$,R) coordinates to write Einstein field
equations and perform the imbedding by the energy-momentum tensor 
(for details see \cite{Gautreau}):

\begin{eqnarray*}
ds^{2}=d\tau^{2}-[dR-(1-A)^{1/2}d\tau]^{2}-R^{2}d\Omega^{2}, \\
T_{\mu\nu}=T_{\mu\nu}(cosmic\ fluid)+T_{\mu\nu}(Schwarzschild\ mass).
\end{eqnarray*}

The complete modification of the Schwarzschild geometry within cosmic mean mass-density 
$\rho_{m,0}$, 
the cosmological constant $\rho_{\Lambda}$ 
\cite{Gautreau} and with the relic cosmic acceleration becomes (Eqs. (5) and (6)):

\begin{eqnarray}
A(T,R)=\frac{B(T,R)}{\tau_{,T}^{2}}=
1-2G_{N}M/R-H_{0}^{2}R^{2}(\Omega_{m}+\Omega_{\Lambda})+2\sqrt{\Sigma}H_{0}R, \\
\rho_{m,0}=\Omega_{m}\rho_{c,0},\ \rho_{\Lambda}=
\Omega_{\Lambda}\rho_{c,0},\ G_{N}\rho_{c,0}=\frac{3}{8\pi}H_{0}^{2}.
\end{eqnarray}

We employ the fact that for small R we have $\tau_{,T}\rightarrow 1$, and for large R
the Einstein-de Sitter model ($\Omega_{\Lambda}=0$) leads to \cite{Gautreau}:

\begin{eqnarray}
T=\tau [1+\frac{1}{2}(2R/3\tau )^{2}]^{3/2}, \hspace{60 mm} \nonumber \\
\frac{\partial \tau}{\partial T}\mid_{today}=1+\frac{1}{3}(\frac{R}{\tau_{U}})^{2}+...
=1+{\cal O}(10^{-28}),\ \tau_{U}\simeq H_{0}^{-1}, \ R\simeq 10^{12}m.
\end{eqnarray}

The inclusion of the positive or negative cosmological constant can change our
estimate of the small correction only for, roughly, one order of magnitude
\cite{Weinberg}, thus confirming our ignorance
of any correction to $\tau_{,T}$.

Let us make numerical evaluations and comparisons of the cosmological contributions to the 
Schwarzschild metric relevant for solar system dynamics.
We denote Hubble's constant by $H_{0}$, the solar mass by $M_{\odot}$,
the typical planetary distance from the Sun by $\bar{R}$; $a_{M}$ is
the Newtonian acceleration of the Sun, $a_{\Sigma}$ is the anomalous acceleration
of the solar system
caused by cosmological acceleration, $a_{\rho + \Lambda}$ is the acceleration 
due to the cosmic mass density and the cosmological constant:

\begin{eqnarray}
H_{0}=h_{0}\times 100 km/Mpc/s,\ h_{0}=0.71, \hspace{20 mm} \nonumber \\
M_{\odot}=2\times 10^{33}g,\ \bar{R}=10^{12}m,\hspace{30 mm} \nonumber \\
a_{M}=G_{N}M_{\odot}/\bar{R}^{2},\ a_{\Sigma}=\sqrt{\Sigma}cH_{0}, \hspace{15 mm} \nonumber \\
a_{\rho + \Lambda}=-\bar{R}H_{0}^{2},\ \Omega_{m}+\Omega_{\Lambda}=1,\ \Sigma=10^{-3}, 
\nonumber \\
a_{\Sigma}/a_{M}={\cal O}(10^{-7}),\ a_{\rho + \Lambda}/a_{\Sigma}={\cal O}(10^{-13}), 
\nonumber \\
(\frac{G_{N}M_{\odot}}{\bar{R}})^{2}=2.22\times 10^{-18} <
\sqrt{\Sigma}H_{0}\bar{R}=2.42\times 10^{-16}.
\end{eqnarray}

We see that the cosmic mean-mass density and the cosmological constant contributions to
the solar system dynamics are negligible. The cosmic relic acceleration term
could be important for a precise PPN analysis appearing as a constant effective
acceleration towards the Sun comparable or greater than the second order PPN terms
\cite{LATOR}. The LATOR mission has sensitivity large enough  to measure 
the effect. The Pioneer 10 and 11 anomalous acceleration acts with the same
sign and R-scaling as the cosmic acceleration relic, but it is more than one order
of magnitude larger \cite{Pioneer}:

\begin{eqnarray}
a_{\Sigma}=\sqrt{\Sigma}cH_{0}=0.22\times 10^{-10} m s^{-2}\ for\ 
\Sigma=10^{-3}, \\
a(Pioneer)=8.74\times 10^{-10} m s^{-2}. \hspace{30 mm}
\end{eqnarray}

The large acceleration parameter $\Sigma={\cal O}(1)$ is cosmologically forbidden 
\cite{Palle2,Palle3}.
The Pioneer anomalous acceleration is not consistent with planetary 
orbits data (see the second paper of ref. (8) for details).
Thus, only forthcoming precise observations could resolve the puzzle of the constant anomalous 
universal acceleration 
in the solar system.


\begin{thebibliography}{300}
\bibitem{Gautreau} 
 R. Gautreau:
   Phys. Rev. {\bf D 29} (1984) 186;
   Phys. Rev. {\bf D 29} (1984) 198
\bibitem{Palle1} 
   D. Palle:
   {\bf astro-ph/0407122}, to appear in Nuovo Cimento B
\bibitem{Palle2}
   D. Palle:
   {\bf astro-ph/0312308}
\bibitem{Palle3}
   D. Palle:
   Nuovo Cimento {\bf B 111} (1996)  671
\bibitem{Palle4}
   D. Palle:
   Nuovo Cimento {\bf B 114} (1999)  853
\bibitem{Weinberg} 
   S. Weinberg:
   {\em Gravitation and cosmology},
   John Wiley and Sons, New York, 1972
\bibitem{LATOR} 
   S. G. Turyshev , M. Shao and K. L. Nortvedt Jr.:
   {\bf gr-qc/0401063}
\bibitem{Pioneer}
   J. D. Anderson et al:
   Phys. Rev. Lett.{\bf 81} (1998) 2858;
   Phys. Rev. {\bf D 65} (2002) 082004 
\end{thebibliography}
\end{document}